\documentclass[fleqn,usenatbib]{mnras}
\usepackage{newtxtext,newtxmath}
\usepackage[T1]{fontenc}
\DeclareRobustCommand{\VAN}[3]{#2}
\let\VANthebibliography\thebibliography
\def\thebibliography{\DeclareRobustCommand{\VAN}[3]{##3}\VANthebibliography}
\usepackage{graphicx}
\usepackage{amsmath}
\usepackage[latin1]{inputenc}
\usepackage{tikz}
\usetikzlibrary{shapes,arrows}
\usepackage{siunitx} 
\newcommand{\daa}{\Delta\alpha/\alpha}

\newcommand{\fev}{{Fe\sc\,v}}
\newcommand{\niv}{{Ni\sc\,v}}

\title[Continuum fitting white dwarfs]{Measuring the fine structure constant on white dwarf surfaces; uncertainties from continuum placement variations}

\author[C.C. Lee et al]{Chung-Chi Lee$^1$,
John K. Webb$^1$,
Darren Dougan$^2$,
Vladimir A. Dzuba$^3$,
Victor V. Flambaum$^3$.
\\\\
$^1$Clare Hall, University of Cambridge, Herschel Rd., Cambridge CB3 9AL, UK.\\
$^2$Big Questions Institute, Level 4, 55 Holt St., Surry Hills, Sydney, NSW 2010, Australia.\\
$^3$School of Physics, University of New South Wales, Sydney, NSW 2052, Australia.
}
\date{Accepted \phantom{mmmmmmm}. Received \phantom{mmmmmmm}; in original form \phantom{mmmmmmm}}

\pubyear{2022}

\usetikzlibrary{arrows.meta,chains,positioning,quotes,shapes.geometric}

\tikzset{FlowChart2/.style={
startstop/.style = {rectangle, rounded corners, draw, fill=yellow!10,
                    minimum width=3cm, minimum height=1cm, align=center, thick,
                    on chain, join=by arrow},
  process/.style = {rectangle, rounded corners, draw, fill=green!10,
                    text width=4cm, minimum height=1cm, align=center, thick,
                    on chain, join=by arrow},
 decision/.style = {diamond, aspect=1.5, draw, fill=orange!10,
                    minimum width=3cm, minimum height=1cm, align=center, thick,
                    on chain, join=by arrow},
       io/.style = {trapezium, trapezium stretches body, 
                    trapezium left angle=70, trapezium right angle=110, thick,
                    draw, fill=blue!30,
                    minimum width=3cm, minimum height=1cm,
                    text width =\pgfkeysvalueof{/pgf/minimum width}-2*\pgfkeysvalueof{/pgf/inner xsep},
                    align=center,
                    on chain, join=by arrow},
    arrow/.style = {thick,-latex'}
                        }
        }

\begin{document}
\label{firstpage}
\pagerange{\pageref{firstpage}--\pageref{lastpage}}
\maketitle

\begin{abstract}
Searches for variations of fundamental constants require accurate measurement errors. There are several potential sources of errors and quantifying each one accurately is essential. This paper addresses one source of uncertainty relating to measuring the fine structure constant on white dwarf surfaces. Detailed modelling of photospheric absorption lines requires knowing the underlying spectral continuum level. Here we describe the development of a fully automated, objective, and reproducible continuum estimation method, based on fitting cubic splines to carefully selected data regions. Example fits to the Hubble Space Telescope spectrum of the white dwarf G191-B2B are given. We carry out measurements of the fine structure constant using two continuum models. The results show that continuum placement variations result in small systematic shifts in the centroids of narrow photospheric absorption lines which impact significantly on fine structure constant measurements. This effect must therefore be included in the overall error budget of future measurements. Our results also suggest that continuum placement variations should be investigated in other contexts, including fine structure constant measurements in stars other than white dwarfs, quasar absorption line measurements of the fine structure constant, and quasar measurements of cosmological redshift drift.
\end{abstract}


\section{Introduction}

Modelling absorption features in high-resolution spectra requires an estimate of the unabsorbed continuum level, since the optical depth $\tau_{\lambda} = -\ln(I_{\lambda}/I_{0,\lambda})$ (where $I_{\lambda}$ is the observed spectral intensity and $I_{0,\lambda}$ is the unabsorbed continuum level); one must know $I_{0,\lambda}$ to fit a theoretical model of $\tau_{\lambda}$ to observational data. Broadly, there are three options: (1) estimate $I_{0,\lambda}$ simultaneously with estimating the theoretical $\tau_{\lambda}$, (2) estimate $\tau_{\lambda}$ and $I_{0,\lambda}$ independently, or (3) make a preliminary estimate of $\tau_{\lambda}$ and $I_{0,\lambda}$ and subsequently simultaneously refine the parameters for $I_{0,\lambda}$ whilst minimising $\chi^2$ to model $\tau_{\lambda}$. (1) can be applied but only in limited circumstances: the continuum function must be very simple, otherwise it is easy to create severe degeneracy between $I_{0,\lambda}$ and $\tau_{\lambda}$ within the fitting region (essentially the fitting function is limited to be no more complex than a straight line segment, which in turn restricts the applicability to very small spectral segments). In this context, degeneracy is  particularly likely during a non-linear least squares modelling procedure when, in the early iterations, $\tau_{\lambda}$ can be far from the best-fit solution. (2) can also be used but then the uncertainty estimates for the final absorption line parameters do not reflect continuum uncertainties. (3) helps to alleviate the shortfallings encountered with the first two methods and is the preferred method. A further motivation for (3) is that it is generally beneficial to already have a reasonably good estimate of $I_{0,\lambda}$ prior to detailed modelling of $\tau_{\lambda}$. 

Previous works describing continuum fitting methods include \cite{Suzuki2005}, who use principal component analysis methods to estimate continua in high resolution Lyman-$\alpha$ forest data (noting where PCA methods are prone to fail). \cite{Lee2012,Davies2018} discuss principal component analysis methods to estimate continua in the Lyman-$\alpha$ forest, for SDSS data i.e. at lower spectral resolution. \cite{Sanchez2018} describe their Python software {\sc statcont}. \cite{Ciollaro2014} develop methods applied to Hubble Space Telescope Faint Object Spectrograph and Baryon Oscillation Spectroscopic Survey spectra. Despite of all these previous works, after extensive searching and experimenting, none had the level of automation, reproducibility, and accuracy required. We expand on this point in Section \ref{sec:G191}. The new continuum-fitting method described in this paper helps combat these effects.

Another driver for the work described in this paper (and its timeliness) concerns the increasing availability of high quality data using new ground-based instruments. The ESPRESSO spectrograph on the European Southern Observatory's VLT is designed to make the next advance in quasar spectroscopy and the forthcoming ELT with the HIRES spectrograph, the advance after that. Varying fundamental constants is one of the drivers for ESPRESSO and one of the prime science ELT goals e.g. \cite{Marconi2016}. As data quality improves, so does the need for optimal analysis tools. However, we defer a detailed discussion relating to quasar spectroscopy; measurements of $\daa$ from quasar absorption systems depend sensitively on the velocity structure in the absorbing medium. In turn, the velocity structure returned from a fitting procedures like {\sc vpfit} \citep{ascl:VPFIT2014, web:VPFIT, WebbVPFIT2021} and {\sc ai-vpfit} \citep{Lee2020AI-VPFIT} is non-unique \citep{Lee2021} and is likely to depend on the local continuum estimate. The same is not true for the single-redshift photospheric absorption lines seen in white dwarf spectra \citep{Preval2013}. Therefore, the means of quantifying the impact of local continuum uncertainty for quasar measurements is very different to the white dwarf case. For this reason, we restrict the discussion in the present study to white dwarfs.

In this paper we describe a fully automated continuum fitting method, based on fitting cubic splines to the data, after removing regions containing spectral features (Section \ref{sec:Splines}). Example fits are illustrated in Section \ref{sec:examples}. Having more than one continuum model is enables us to estimate the contribution of continuum placement uncertainty to the overall fine structure constant uncertainty. We examine this point in Section \ref{sec:impact}.

\section{Astronomical data} \label{sec:G191}

The astronomical data used is the Hubble Space Telescope STIS spectrum of the well known white dwarf G191-B2B. The data were obtained using the E140H grating, with resolution $R\approx114,000$ (the highest currently available for UV astronomical spectroscopy), and have a signal to noise $\sim$100 per resolution element. A comprehensive description covering both the observational data and the data reduction processes is given in Appendix A of \cite{Hu2019}. This object was chosen because: (a) several detailed studies already exist, including measurements of $\daa$ \citep{Berengut2013, Hu2021}; (b) the STIS echelle instrument format means individual spectral orders are extracted from the detector individually before combination to form a continuous 1-d spectrum. This means the data are fairly ``challenging'' in that weak undulations in the apparent continuum are present due to imperfect order flattening prior to combination (as seen in the illustrations later in this paper); (c) the spectrum is riddled with hundreds of photospheric absorption lines \citep{Preval2013} that must be detected and removed prior to (or in conjunction with) estimating the underlying continuum. Nevertheless, the G140H STIS spectrum of G191-B2B may be considered atypical due to its high signal to noise such that the settings used in the continuum fitting code (Section \ref{sec:Splines}) for this particular spectrum may need adjustment in other applications.

\section{Atomic data}

In order to assess the impact of variations in the fitted continuum level on $\daa$ measurements (in this paper, for white dwarf spectra -- see Section \ref{sec:impact}), we broadly follow the procedures described in \cite{Berengut2013}. To do so we require input rest-frame wavelengths and sensitivity coefficients describing wavelength shifts as a function of $\daa$ (q-coefficients), calculated using
\begin{equation}
    \omega_z = \omega_0 + q \left(\frac{\alpha_z^2}{\alpha_0^2} - 1 \right)
\end{equation}
where $\omega$ denotes transition frequency and the subscripts $z$ and $0$ indicate redshifted and terrestrial values. Calculations of sensitivity coefficients applied to an astronomical context were first described in \cite{Dzuba1999a, Dzuba1999b}. {\niv} (and {\fev}) q-coefficients for UV transitions were first reported in \cite{Ong2013}. Improvements to the \cite{Ong2013} calculations are provided in \cite{Hu2021} but only for {\fev} and not for {\niv}. For the calculations described in the present paper, we have revised the {\niv} q-coefficients given in \cite{Ong2013} using the same methods described in \cite{Hu2021}. We therefore do not re-iterate technical details provided in those papers and provide the relevant atomic data tables as online supplementary material, in which the transition frequencies and  q-coefficients are presented in units of cm$^{-1}$.

Whilst in this analysis we opted to use {\niv}, we could have equally used {\fev} but chose {\niv} purely because a detailed analysis of that species (in the spectrum of G191-B2B) is underway, to be published separately. {\niv} wavelengths and energy level measurements have been published in \cite{Raassen1976a, Raassen1976b, Ward2019}. However, errors in the published values were found and corrected by \cite{NIST_ASD}. The {\niv} wavelengths used in this analysis are those given in the {\it NIST Atomic Spectra Database database v10}.

\section{Cubic splines with line removal} \label{sec:Splines}

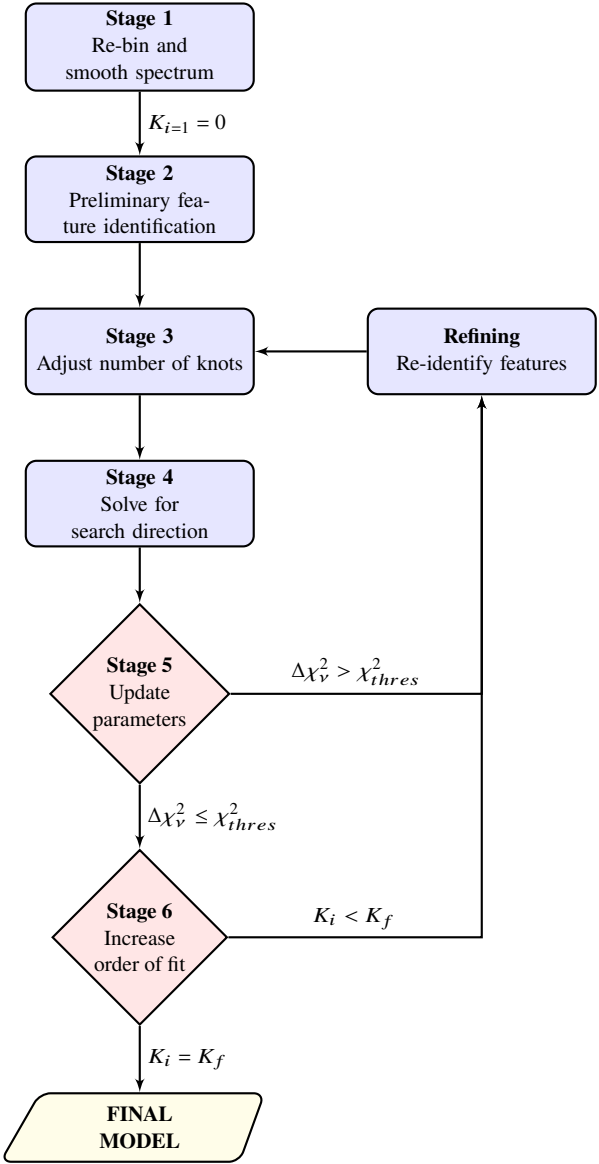
\begin{figure}
\centering
\tikzstyle{decision} = [diamond, draw, thick, fill=red!10, text width=5em, text badly centered, node distance=3cm, inner sep=0pt]
\tikzstyle{block} = [rectangle, draw, thick, fill=blue!10, text width=10em, text centered, rounded corners, minimum height=4em]
\tikzstyle{line} = [draw, thick, -latex']
\tikzstyle{cloud} = [draw, ellipse,fill=red!20, node distance=3cm, minimum height=2em]
\tikzstyle{end} = [trapezium, draw, thick, trapezium right angle=110, text width=7em, text centered, rounded corners, fill=yellow!10, node distance=1.9cm, minimum height=3.2em]
\begin{tikzpicture}[node distance = 2.5cm, auto]
\node [block] (stage1) {{\bf Stage 1} \\ Re-bin and smooth spectrum};
\node [block, below of=stage1, node distance=2.0cm] (stage2) {{\bf Stage 2} \\ Preliminary feature identification};
\node [block, below of=stage2, node distance=2.0cm] (stage3) {{\bf Stage 3} \\ Adjust number of knots};
\node [block, right of=stage3, node distance=4.5cm] (refining) {{\bf Refining} \\ Re-identify features};
\node [block, below of=stage3, node distance=2.0cm] (stage4) {{\bf Stage 4} \\ Solve for search direction};
\node [decision, below of=stage4, node distance=2.5cm] (stage5) {{\bf Stage 5} \\ Update parameters};
\node [decision, below of=stage5, node distance=3.2cm] (stage6) {{\bf Stage 6} \\ Increase order of fit};
\node [end, below of=stage6, node distance=2.5cm] (final) {{\bf FINAL MODEL}};
\path [line] (stage1) -- node[right]{$K_{i=1}=0$}(stage2);
\path [line] (stage2) -- (stage3);
\path [line] (refining) -- (stage3);
\path [line] (stage3) -- (stage4);
\path [line] (stage4) -- (stage5);
\path [line] (stage5) -| node[near start,above]{$\Delta \chi^2_\nu > \chi^2_{thres}$} (refining);
\path [line] (stage5) -- node[right]{$\Delta \chi^2_\nu \le \chi^2_{thres}$} (stage6);
\path [line] (stage6) -| node[near start,above]{$K_i<K_f$} (refining);
\path [line] (stage6) -- node[right]{$K_i=K_f$} (final);
\end{tikzpicture}
\caption{This flow chart summarises the continuum fitting model based on spline fitting with automated line removal. $K_i$ indicates the number of cubic spline knots used at the $i^{th}$ iteration. $K_f$ is the number of knots used for the final best-fit continuum model. See Section \ref{sec:Splines} for details.
\label{fig:flowchart}
}
\end{figure}

The first step towards obtaining the final estimated continuum is to make a preliminary model using cubic splines. To do this we first identify significant absorption and emission profiles in order to be able to fit cubic splines to feature-free spectral segments. The method for identifying features is discussed shortly. One problem to avoid is allowing the continuum model to follow an absorption profile, thereby effectively removing it. For example, consider a shallow absorption profile, spread across $\sim$2 {\AA} such that it is reasonably well fitted by a cubic spline with a knot interval of 0.5{\AA}, such that the absorption line can be accidentally removed. This problem can be avoided by flagging the data points within the absorption line prior to fitting the cubic spine model. Doing this interactively prior to the continuum fitting stage would be both time consuming and subjective. Therefore, we instead iteratively increase the spline order: the initial model is a single cubic spline, after which the number of knots is doubled at each subsequent iteration. Figure \ref{fig:flowchart} shows two loops: loop A (iterating over stages 2-5) and loop B (iterating over stages 2-6). Each loop has its own specific stopping criterion. These are discussed shortly (Sections \ref{sec:stg5} and \ref{sec:stg6}).

\subsection{Stage 1: Re-bin and smooth spectrum}
\label{sec:stage1}

The optical depth associated with absorption features in the spectrum is given by
\begin{equation}
\tau_{tot}(\lambda) = \sum_i \tau_{abs}^i(\lambda)
\end{equation}
where $\tau_{abs}^i$ is the optical depth associated with an individual $i^{th}$ absorption line and where the sum is taken over all absorption (or emission) lines present in the model. The observed spectrum is the sum of three terms,
\begin{equation}
\label{eq:I_tot}
    I(\lambda) = I_{0}(\lambda) e^{-\tau_{tot}(\lambda)} + I_{em}(\lambda) + I_{n}(\lambda) \,,
\end{equation}
where $I_{0}$, $I_{em}$, and $I_{n}$ are the underlying continuum from the object being observed, the emission line spectrum, and noise, respectively. The noise term has several contributions, e.g. photon counting from the object, photon counting from the sky, assuming the spectrum is sky-subtracted, detector read-out noise, detector dark current, cosmic rays, and any detector defects that have not been fully removed in the data pre-processing. 

In Stage 1 we smooth the data by convolving the spectrum with a Gaussian function. We first re-bin the spectrum onto a finer grid (with pixel size 1/10 the original value). The re-binned spectrum is then smoothed using a Gaussian. The FWHM of the smoothing Gaussian is taken to be a multiple of the mean original pixel size $\bar{x}$ (over the entire spectrum). The default pixel space is in {\AA} and in practice a value of 3$\bar{x}$ for the FWHM was found to work well, but this is a user-defined quantity. The observed pixel size is approximately constant in velocity units, so convolution is done in velocity space. This initial smoothing is helpful in the next Stage where we calculate the derivative of the spectrum. Stage 1 is only carried out once - it is not included in subsequent loops (as illustrated in Fig.\ref{fig:flowchart}).

\subsection{Stage 2: Preliminary feature identification}
\label{sec:stage2}

In order to determine whether a pixel in the spectrum falls within an absorption (or emission) feature, or whether it can be considered as a continuum pixel, two properties are examined: the closeness of the pixel intensity to the current continuum estimate and the derivative of the smoothed spectrum. The derivative spectrum is obtained using the re-binned smoothed spectrum from Stage 1. 

The first step is to make a preliminary identification of all absorption (or emission) features in the data. To do this we make simple use of the rebinned, smoothed derivative spectrum from Stage 1; the largest $x$\% of the data points in that spectrum (across the entire spectrum) are identified (using a simple array sort) and excluded . In practice we found $x=30$\% works well. This process of course identifies line edges where the derivative is high (but also, necessarily) line centres are {\it not} selected.

Having selected the points as described above, an initial straight line continuum is fitted. This initial continuum will of course be only a very crude representation of the true continuum. Then, now having selected low-slope data points (the remaining $100-x$\%) and with a preliminary (lowest order) continuum fit, we calculate the normalised chi-squared, $\chi^2_\nu$ over the selected data points, 
\begin{align}
&\chi^2_\nu \approx \frac{\chi^2}{M_s} \label{eq:chisqn} \\
&\chi^2 = \sum_{i=1}^{M_s} \left( \frac{I_i - C_i}{\sigma_i} \right)^2 \,, \label{eq:chisq}
\end{align}
where $C_i$ is the current estimated continuum value at $i^{th}$ pixel, $\sigma_i$ is the 1-$\sigma$ intensity uncertainty, and $M_s$ is the total number of selected data points. Prior to refinement, with only a crude continuum fit, the initial value of $\chi_\nu^2 \gg 1$. To reject data points at line centres we search for pixel values having $(I_i - C_i)^2/\sigma_i^2 > \zeta^2 \chi^2_\nu$, using a default value of $\zeta=3$ to correspond approximately to $3\sigma$ deviations.

\subsection{Stage 3: Adjust number of knots}
\label{sec:stage3}

At first pass through Stage 3 (when $K_i=0$), the continuum model is very simple (initially only a straight line) so Stage 3 applies only to higher iterations. During the development of a continuum model, the number of cubic splines increases. It is possible that at certain positions along the spectrum, only a few continuum data points are available to fit. If this happens, that segment of the continuum will of course be poorly determined. Therefore, Stage 3 checks that a sufficiently large number of data points is present between each knot pair. The minimum acceptable number of continuum data points between each knot pair is a user-defined parameter, having a default value of $k_{merge}=1/3$, this parameter being the fraction of pixels available within the range defined by the knot pair. When there are too few continuum data points left between any particular knot pair, the two flanking regions are checked and the test region merged into whichever side region has the least continuum data points (such that the knot spacing is no longer necessarily regular).

\subsection{Stage 4: Solve for search direction}

By the end of Stage 3, the model continuum can comprise a large number of free parameters. The goal is to minimise $\chi^2$ between the cubic spline model and the set of continuum data points. We do this using a standard non-linear least-squares procedure. The components of the gradient vector $g(a,b)$ and of the Hessian matrix $H_{ab}$ are computed using first-order numerical finite-difference derivatives,
\begin{align}
&g_a = \frac{d \chi^2}{dy_a} \,, \\
&H_{ab} = \frac{d^2 \chi^2}{ dy_a dy_b} \,,
\end{align}
where $y_a$, $y_b$ are the knot values (i.e. intensities) at positions $a$ and $b$. Then we apply a standard Gauss-Newton minimisation procedure,
\begin{equation}
    H_{ab} p_b = - g_a \,,
\end{equation}
where $p_b$ is the search direction, providing the best-fit set of knot intensities. Detailed descriptions of optimisation methods, including this one, can be found in, for example, \cite{GMW81}.

\subsection{Stage 5: Update parameters}
\label{sec:stg5}

The model continuum parameters are updated (using univariate minimisation) by finding the scalar $d$ which minimises $\chi^2$. The updated parameters are then
\begin{equation}
    \bar{y}_b = y_b + p_b \times d \,.
\end{equation}
The values of $\chi_\nu^2$ for the current and previous iterations are compared and if they differ by more than a value of $\chi_{thres}^2$, the algorithm returns to an earlier stage (analogous to Stage 2) and re-identifies all absorption and emission features using the newly updated continuum model. This is illustrated in Figure \ref{fig:flowchart} by the ``Refining'' box. If the difference is smaller than $\chi_{thres}^2$, the algorithm moves to Stage 6. The default value of $\chi_{thres}^2$ is $10^{-3}$ (although this can be user-defined). Final results are rather insensitive to this parameter.

\subsection{Stage 6: Increase order of fit}
\label{sec:stg6}

At this stage, the number of cubic spline knots is $K_i$. $K_f$ is the full data range divided by the smallest permitted knot spacing and is a user input, specified to terminate the algorithm. If $K_i < K_f$, the algorithm thus iterates further as indicated in Figure \ref{fig:flowchart}. We choose the form for the evolution of the number of knots $K_i \rightarrow K_{i+1}$ to be
\begin{equation}
    K_{i+1} = \mathrm{Int} \left( \frac{K_f}{2^{n_i}} \right) \,,
\end{equation}
where $n_{i} = \mathrm{Int} (\log_2 K_f -i+1)$, such that the number of knots approximately doubles at each successive iteration.

However, the number of knots used has to be carefully considered. For example, to avoid over-fitting, one should avoid making the knot spacing comparable to or smaller than the width of an absorption line or blended feature. Also, the STIS echelle format means that many spectral orders are pieced together to form a final one dimensional spectrum. The order flattening process is imperfect and small wiggles may remain in the final one dimensional spectrum on scales corresponding to the order separation. Approximately matching the knot density to that scale enables the continuum model to follow these undulations and hence can help remove this effect. Detailed descriptions of STIS data reduction procedures are given in e.g. \cite{Ayres2010, Ayres2022}.

\subsection{Refining: Re-identify Absorption and Emission Lines}\label{sec:refining}

If the conditions for further iterations specified in Stages 5 and 6 are satisfied (see Figure \ref{fig:flowchart}), the algorithm return to re-identify all absorption and emission features, relative to the current continuum fit. The procedure during refinement is similar to Stage 2: 

\begin{enumerate}

\item Moving out from an absorption line centre, one can specify a point at which the observed intensity recovers to the underlying continuum level (within some tolerance). Also, once the intensity recovers to the local continuum level, noise fluctuations mean that the derivative is likely to change sign away from the absorption line centre. Thus we locate the point at which the derivative of the smoothed spectrum changes sign, moving outwards from the line centre. These points (left and right of each spectral line centre) provide a {\it preliminary} estimate of the feature boundaries and hence define an initial set of continuum pixels. However, several further steps refine these initial estimates.

\item Having identified feature edges, i.e. discarded pixels within and near to absorption (or emission) features, we now attempt to put previously excluded pixels back into the set used for continuum fitting. For each pixel in the unsmoothed spectrum, we calculate the intensity difference between it and the current continuum estimate. If the pixel in the unsmoothed spectrum is very close to the current continuum, it can be ``re-assigned'' as a pixel to be used the continuum fit, whether or not it formed part of the preliminary continuum pixel set. However, to be re-assigned, it must satisfy a second condition: the derivative of the smoothed spectrum at that point must be below a specified threshold. In this way, additional pixels are picked up to be used in fitting a continuum function.

To express the above more rigorously, let the intensity of the $i^{th}$ pixel in the original unsmoothed spectrum be $I_i$, with uncertainty $\sigma_i$. Also let the intensity of the $i^{th}$ pixel in the current continuum estimate be $C_i$. Then let
\begin{equation}
 \Delta_i = \lvert (I_i - C_i)/\sigma_i \rvert,
\end{equation}
and $d_i$ be the derivative of the smoothed original spectrum. We then require both $\Delta_i$ and $d_i$ to be smaller than threshold values. The threshold values are defined as follows. For all continuum pixels, we compute the distribution of $\Delta_i$ and determine its standard deviation, $\sigma_D$. The same is done for $d_i$ to determine $\sigma_d$. For a pixel to be ``re-assigned'' as a continuum pixel, we then require, simultaneously,
\begin{equation}
\Delta_i < \sigma_D \,\,\, \textrm{and} \,\,\, d_i < \sigma_d
\end{equation}

\item A further check is now made on pixels that are deemed to be free from absorption or emission features i.e. pixels that are used to fit a continuum fit to. For this test, we use a normalised spectrum i.e. the original spectrum has been divided by the continuum fit. For each contiguous continuum segment i.e. a set of $M$ pixels that are all deemed to be continuum pixels, linear regression is used to determine that segment's slope $s$ and its uncertainty $\sigma_s$. Provided $s- k \sigma_s \le 0 \le s+ k \sigma_s$, that segment remains identified as a continuum segment, where $k$ is a user defined parameter with a default value of 4.0.
\end{enumerate}

The first two conditions above generally succeed in identifying and excluding strong and relatively narrow absorption lines whilst the third condition helps to identify shallow and weaker lines. These conditions above apply equally the emission lines (with a minus sign on the tests). However, at this stage there may nevertheless still be true continuum pixels that have been discarded. Therefore one further test is done:

\begin{enumerate}
\item[(iv)] Where a contiguous set of $n$ or more pixels (default $n=10$) lies {\it above} the continuum fit, if $|d_i| < \sigma_d$, those pixels are re-assigned as being pixels to include in the continuum fit.
\end{enumerate}

The whole process described above is illustrated as a flowchart in Figure \ref{fig:flowchart}.

\subsection{Example continuum fits -- the white dwarf G191-B2B} \label{sec:examples}

\begin{figure*}
\centering
\includegraphics[width=0.45\linewidth]{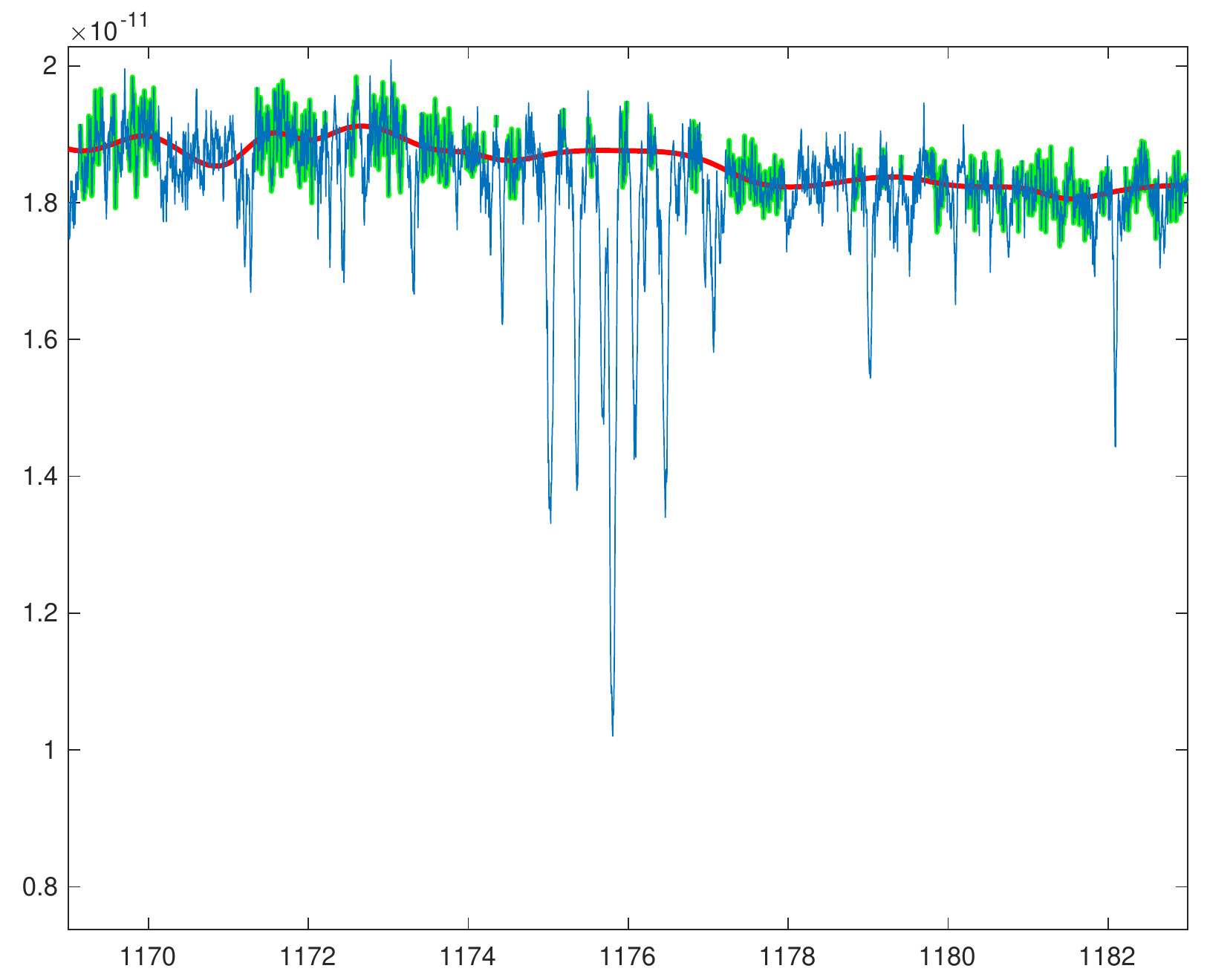}
\includegraphics[width=0.45\linewidth]{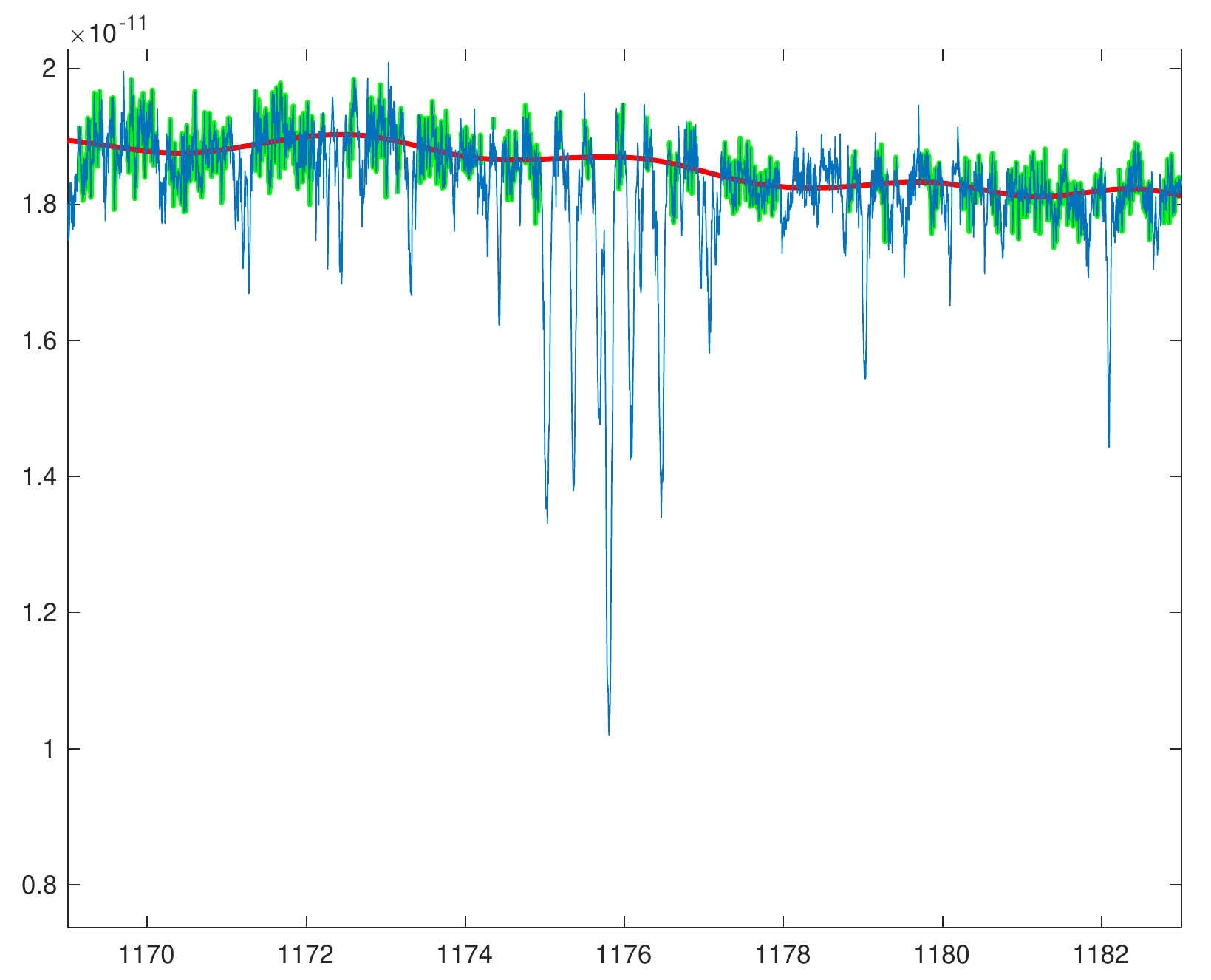} \\
\includegraphics[width=0.45\linewidth]{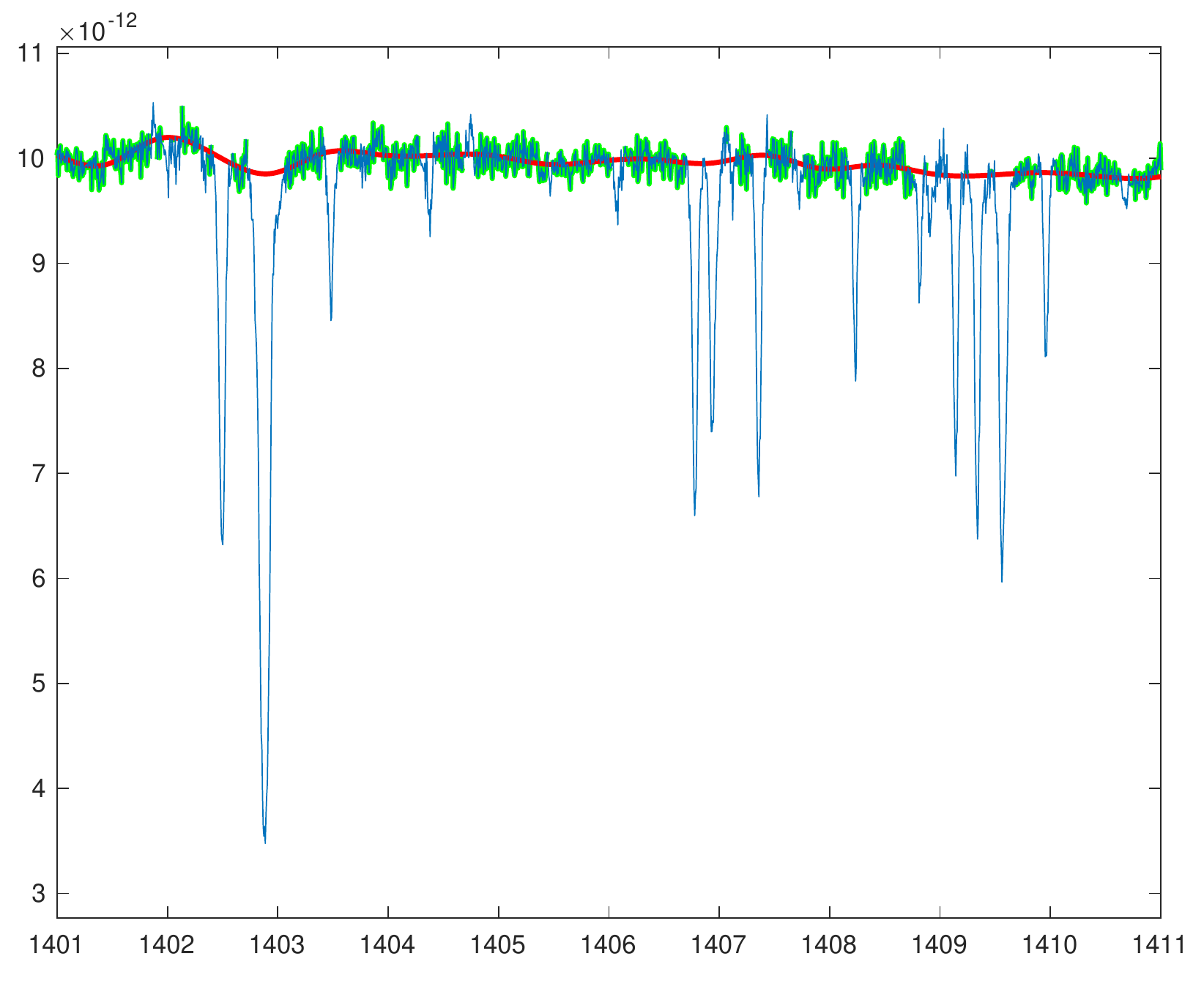}
\includegraphics[width=0.45\linewidth]{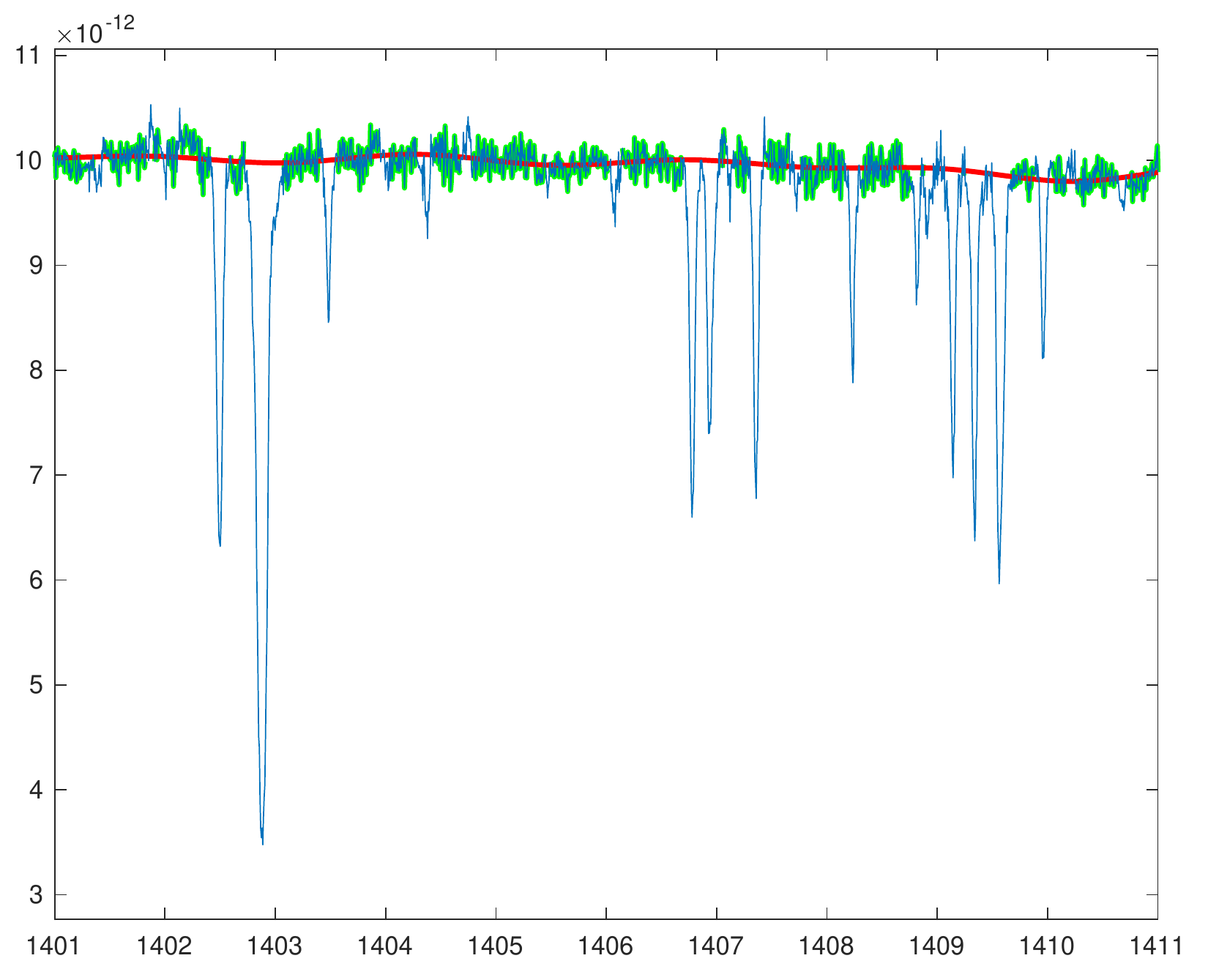} \\
\includegraphics[width=0.45\linewidth]{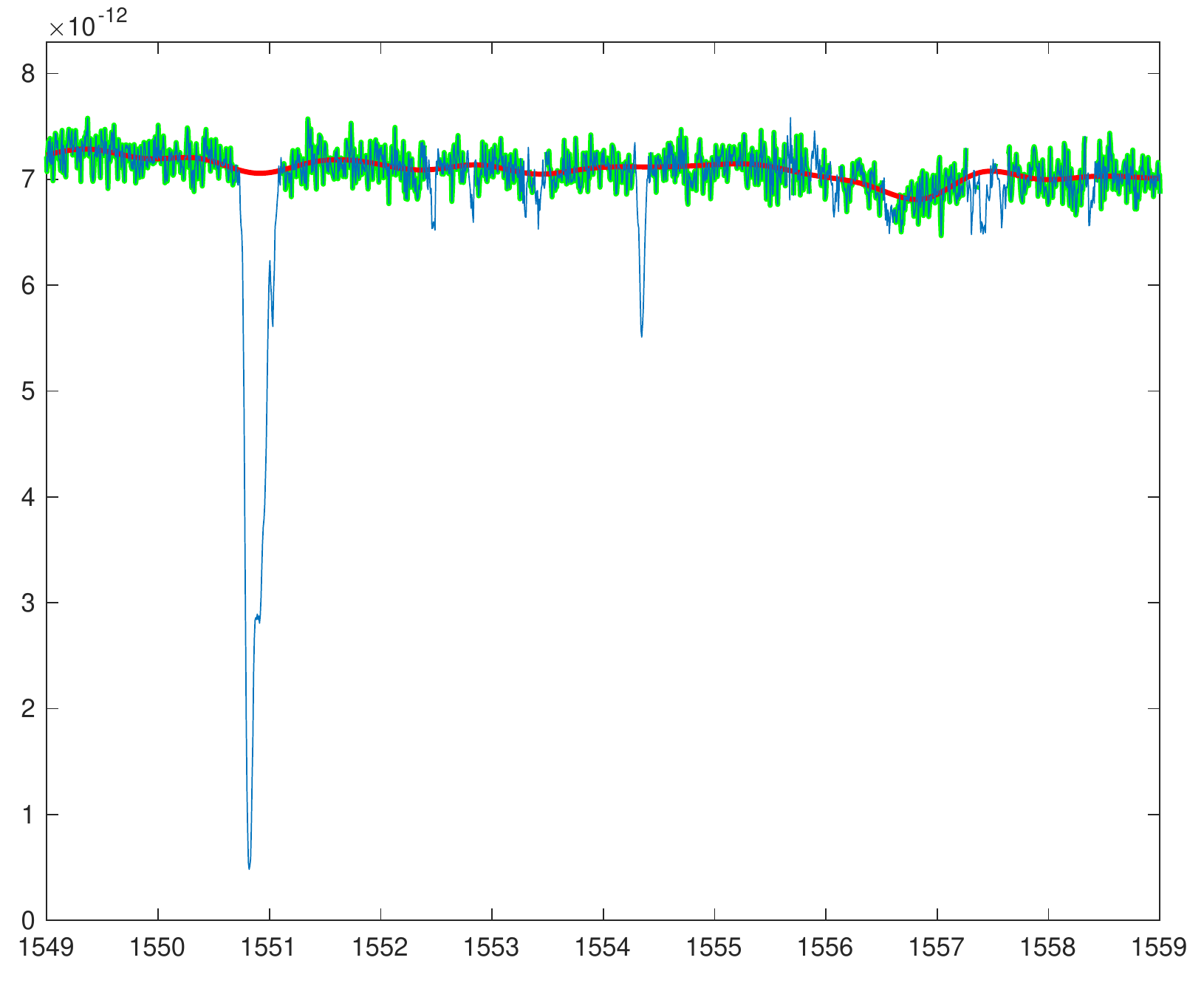}
\includegraphics[width=0.45\linewidth]{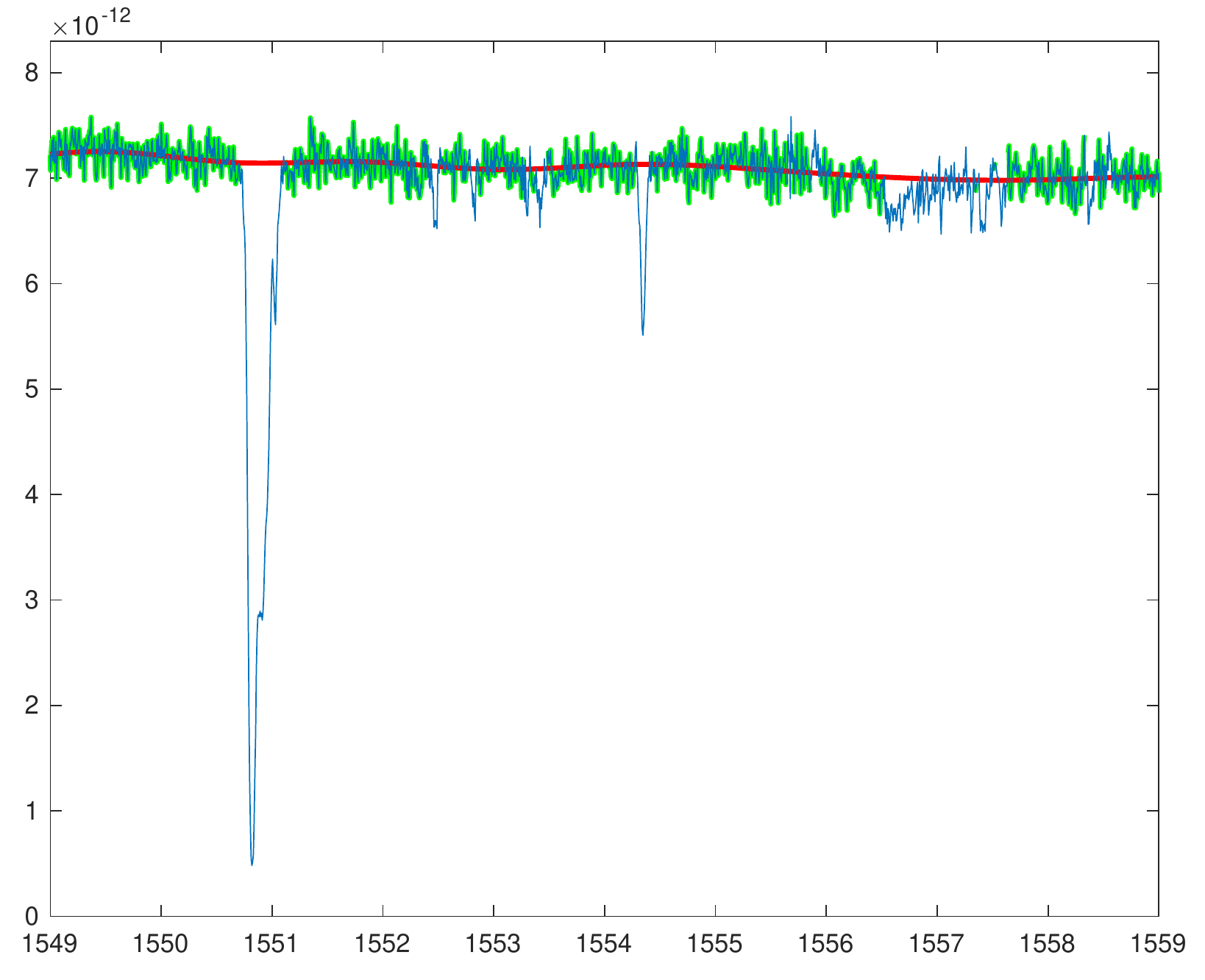}
\caption{Example segments of the HST E140H STIS spectrum of the white dwarf G191-B2B with continuum fits. Left column: 0.5 {\AA} knot spacing, right column: 1.2 {\AA} knot spacing. The red continuous line shows the continuum. The green highlighted regions illustrate pixels designated as line-free regions. The blue continuous line is the original spectrum. The green line shows where the code detected lines and removed those pixels from the continuum fit. See Table \ref{tab:results} for the relevant continuum model settings.
\label{fig:}
}
\end{figure*}

Figure \ref{fig:} shows 3 representative segments in the spectrum of G191-B2B with the continuum model. Two normalised spectra with knot spacing 0.5{\AA} and 1.2{\AA} were input to {\sc rdgen} to identify absorption features at or above a 5$\sigma$ detection threshold.  
Figures \ref{fig:full0.5} and \ref{fig:full1.2} in Appendix \ref{appendix} illustrate the entire spectrum with model continua. Neither models work particularly well over the broad Lyman-$\alpha$ absorption line. We have not attempted to improve this as it evidently needs treating separately. For further information about {\sc rdgen}, see \cite{RDGEN14}.

\section{Impact of continuum uncertainty on $\daa$ measurements} \label{sec:impact}

Measuring the relative positions of a large number of narrow photospheric absorption lines allows us to explore whether the fine structure constant $\alpha$ varies in the presence of strong gravity. For discussions on this and general overviews see e.g. \cite{magueijo02, Barrow2005, flambaum08, Uzan2011, Berengut2013, Bainbridge2017, Hu2021}. Repeating measurements using different continuum models quantifies the sensitivity of the measured $\alpha$ to small variations in the continuum. We therefore generated two model continua, using knot spacings of 0.5 and 1.2{\AA}. Using these two continua we then detect absorption lines which are significant at or above 5$\sigma$. This identifies 1,548 and 1739 absorption features in the range 1150{\AA} to 1897{\AA} for the two knot spacings. Using custom Python code, the observed feature wavelengths were then shifted to the rest-frame using a G191-B2B redshift of 23.8 $\pm$ 0.03 km s$^{-1}$ from \cite{Preval2013}). The rest-frame wavelengths were then matched to recently updated {\niv} laboratory wavelengths (provided as online supplementary material), where a match was accepted for agreement equal to or better than 3$\sigma$, allowing for both observational and laboratory wavelength uncertainties. For each knot spacing, 299 and 316 {\niv} matches were made respectively, of which 258 and 273 have associated q-coefficients.

A linear regression analysis on the matched {\niv} lines, following the procedures used in \cite{Berengut2013}, produced two new measurements of the variation of the fine structure constant: $\daa = -2.3 \pm 1.0 \times 10^{-5}$ for knot spacing of 0.5 {\AA} and $\daa = -1.9 \pm 1.0 \times 10^{-5}$ for knot spacing of 1.2{\AA}. The statistical uncertainty of $1.0 \times 10^{-5}$ is from standard linear least squares. The normalised $\chi^2$ was 2.1 for each knot spacing, where $\chi^2$ is the usual sum of the squares of $k$ independent standard normal random variables. Clearly 2.1 is above the canonical value of unity for a good fit to the data, but this is neither surprising nor avoidable: (i) there will undoubtedly be a large number of weak unidentified absorption lines falling below the 5$\sigma$ detection threshold used, adding to the observed scatter in the regions we have taken to be continuum; (ii) other instrumental and data processing artefacts inevitably remain in the data that may not be fully modelled by the continuum fits (e.g. imperfect flattening of individual spectral orders prior to forming a final one-dimensional spectrum). The empirical pixel to pixel scatter is therefore larger than the formal error estimate which is derived as usual using Gaussian statistics. The results described in this section are summarised in Table \ref{tab:results}. 

\begin{table}
\centering
    \begin{tabular}{|l|r|r|}
      \hline
        \textbf{Knots (\AA)} & \textbf{0.5} & \textbf{1.2} \\
        \hline\hline
        5$\sigma$ detect. & 1548 & 1739 \\
        Matches & 258 & 273 \\
        $\chi^2_{\nu}$ & 2.07 & 2.10 \\
        $\daa$ &	$-2.3 \pm 1.0$ & $-1.9 \pm 1.0$ \\
        \hline
    \end{tabular}
    \caption{The impact of using different white dwarf continuum models on measurements of $\daa$ (in units of $10^{-5}$). Two knot spacings are used. The other parameter settings are: smoothing FWHM $3\bar{x}$ (Section \ref{sec:stage1}), $\zeta=3$ (Section \ref{sec:stage2}), $k_{merge}=1/3$ (Section \ref{sec:stage3}), $n=10$ (Section \ref{sec:refining}). In this example, the additional uncertainty on $\daa$ is 40\% of the statistical uncertainty. Observed and laboratory wavelength matches were made using the newly calculated NiV q-coefficients and the \protect\cite{Ward2019} NiV laboratory wavelengths updated as part of this work by A. Kramida. Both atomic data sets are provided as online supplementary material.
    \label{tab:results}}
\end{table}

\section{Discussion and future work} \label{sec:discussion}

In this paper we have focused on two things: (i) developing a new method to obtain good continuum models for HST/STIS white dwarf spectra, and (ii) examining the impact of continuum placement uncertainties on measurements of the fine structure constant in the strong gravitational fields on white dwarf photospheres. The advantage of the kind of automation described here is of course objectivity and reproducibility. Our most important result is that slight changes in the underlying continuum {\it can} impact significantly on $\daa$ measurements, as illustrated in Table \ref{tab:results}. By ``significantly'', we mean here that the additional uncertainty on $\daa$ associated with the continuum placement level is a non-negligible fraction of the statistical uncertainty on $\daa$ (40\% in the example given -- see Table \ref{tab:results}). 
There is a caveat, or at least a further consideration: the analysis in this paper is based on the linear regression method first emplyed (in this context) in \cite{Berengut2013}. A different approach is that of \cite{Hu2021}, in which absorption lines are modelled using Voigt profiles. In the latter, the fits to local continuum segments can be refined using additional free parameters in the non-linear least squares fit (see \cite{web:VPFIT} for details). That process may substantially reduce the 40\% error contribution described above.

The numerical results we have given are illustrative and specific to the G191-B2B spectrum used i.e. they cannot be applied generally to other spectra and other continuum models. For other spectra, the spectral S/N, the spectral resolution, the wavelength coverage of the data, and hence the line IDs may all be different, and even the species may be different (e.g. {\fev} instead of {\niv}). Therefore continuum placement uncertainties need to be computed explicitly and taken into account in the overall error budget in any future white dwarf $\daa$ measurements.

The work presented here also raises corresponding questions about the impact of continuum placement errors on $\daa$ measurements in quasar absorption systems. That topic is of growing interest, given the technological development of laser frequency combs, the new ESPRESSO spectrograph on the VLT, and the forthcoming ELT; these two facilities are expected to devote a significant effort to quasar $\daa$ measurements e.g. \cite{Marconi2016, ELT2018}. We have nevertheless avoided quasar measurements in this paper because the impact on $\daa$ needs a different kind of analysis than the one described in this paper; one cannot simply re-run the same least-squares fit to obtain $\daa$ for several different input continua. In the case of quasars, the absorption system model itself (i.e. the number of absorption components as well as their parameters) is likely to change, such that one must construct the model from scratch for each different input continuum. Moreover, for analyses involving Voigt profile modelling (as is generally the case for quasar $\daa$ measurements), the most prominent codes ({\sc vpfit} or {\sc ai-vpfit}) are able to solve for a simple linear (local) continuum fit as part of the non-linear least squares minimisation, thereby refining the (already good) fits described in Section \ref{sec:Splines} and illustrated in Section \ref{sec:examples}. All these things require supercomputer calculations and are outside the scope of the present white dwarf study.

Another important ELT science driver is direct measurements of cosmological redshift drift using absorption features in quasar spectra, \cite{Sandage1962} and e.g. \cite{Liske2008}. The degree to which continuum placement uncertainties may impact on those measurements is yet to be explored, as it is possible the effect may be even more significant than we have found for white dwarf measurements.

\section*{Acknowledgements}
We are very grateful to Alexander Kramida and to Tom Ayres for invaluable discussions.

\section*{Data Availability}
Based on observations made with the NASA/ESA Hubble Space Telescope, obtained from the data archive at the Space Telescope Science Institute. The STIS spectra of G191-B2B are available from the Barbara A. Mikulski archive. The continuum fitting code described in this paper, the newly calculated q-coefficients, the {\niv} wavelengths used, and detailed numerical results, are all provided as online supplementary material.

\bibliographystyle{mnras}
\bibliography{contin}

\begin{thebibliography}{}
\makeatletter
\relax
\def\mn@urlcharsother{\let\do\@makeother \do\$\do\&\do\#\do\^\do\_\do\%\do\~}
\def\mn@doi{\begingroup\mn@urlcharsother \@ifnextchar [ {\mn@doi@}
  {\mn@doi@[]}}
\def\mn@doi@[#1]#2{\def\@tempa{#1}\ifx\@tempa\@empty \href
  {http://dx.doi.org/#2} {doi:#2}\else \href {http://dx.doi.org/#2} {#1}\fi
  \endgroup}
\def\mn@eprint#1#2{\mn@eprint@#1:#2::\@nil}
\def\mn@eprint@arXiv#1{\href {http://arxiv.org/abs/#1} {{\tt arXiv:#1}}}
\def\mn@eprint@dblp#1{\href {http://dblp.uni-trier.de/rec/bibtex/#1.xml}
  {dblp:#1}}
\def\mn@eprint@#1:#2:#3:#4\@nil{\def\@tempa {#1}\def\@tempb {#2}\def\@tempc
  {#3}\ifx \@tempc \@empty \let \@tempc \@tempb \let \@tempb \@tempa \fi \ifx
  \@tempb \@empty \def\@tempb {arXiv}\fi \@ifundefined
  {mn@eprint@\@tempb}{\@tempb:\@tempc}{\expandafter \expandafter \csname
  mn@eprint@\@tempb\endcsname \expandafter{\@tempc}}}

\bibitem[\protect\citeauthoryear{{Ayres}}{{Ayres}}{2010}]{Ayres2010}
{Ayres} T.~R.,  2010, In: Proceedings of 2010 Space Telescope Science Institute
  Calibration Workshop ``Hubble after SM4. Preparing JWST'', \href
  {https://ui.adsabs.harvard.edu/abs/2010hstc.workE....D} {p.~7}

\bibitem[\protect\citeauthoryear{Ayres}{Ayres}{2022}]{Ayres2022}
Ayres T.~R.,  2022, \mn@doi [The Astronomical Journal]
  {10.3847/1538-3881/ac3762}, 163, 78

\bibitem[\protect\citeauthoryear{{Bainbridge} et~al.,}{{Bainbridge}
  et~al.}{2017}]{Bainbridge2017}
{Bainbridge} M.,  et~al., 2017, \mn@doi [Universe] {10.3390/universe3020032},
  \href {https://ui.adsabs.harvard.edu/abs/2017Univ....3...32B} {3, 32}

\bibitem[\protect\citeauthoryear{{Barrow}}{{Barrow}}{2005}]{Barrow2005}
{Barrow} J.~D.,  2005, \mn@doi [Philosophical Transactions of the Royal Society
  of London Series A] {10.1098/rsta.2005.1634}, \href
  {https://ui.adsabs.harvard.edu/abs/2005RSPTA.363.2139B} {363, 2139}

\bibitem[\protect\citeauthoryear{Berengut, Flambaum, Ong, Webb, Barrow,
  Barstow, Preval  \& Holberg}{Berengut et~al.}{2013}]{Berengut2013}
Berengut J.~C.,  Flambaum V.~V.,  Ong A.,  Webb J.~K.,  Barrow J.~D.,  Barstow
  M.~A.,  Preval S.~P.,   Holberg J.~B.,  2013, \mn@doi [Phys. Rev. Lett.]
  {10.1103/PhysRevLett.111.010801}, 111, 010801

\bibitem[\protect\citeauthoryear{{Carswell}}{{Carswell}}{2021}]{web:VPFIT}
{Carswell} R.~F.,  2021, Bob Carswell's homepage. Available online:
  \url{https://people.ast.cam.ac.uk/~rfc/} (accessed on 27 January 2022), \url
  {https://people.ast.cam.ac.uk/~rfc/}

\bibitem[\protect\citeauthoryear{{Carswell} \& {Webb}}{{Carswell} \&
  {Webb}}{2014}]{ascl:VPFIT2014}
{Carswell} R.~F.,  {Webb} J.~K.,  2014, {VPFIT: Voigt profile fitting program},
  Astrophysics Source Code Library (\mn@eprint {ascl} {1408.015})

\bibitem[\protect\citeauthoryear{{Carswell}, {Webb}, {Cooke}  \&
  {Irwin}}{{Carswell} et~al.}{2014}]{RDGEN14}
{Carswell} R.~F.,  {Webb} J.~K.,  {Cooke} A.~J.,   {Irwin} M.~J.,  2014, RDGEN,
  Astrophysics Source Code Library, record ascl:1408.017, \href
  {https://ui.adsabs.harvard.edu/abs/2014ascl.soft08017C} {}

\bibitem[\protect\citeauthoryear{{Ciollaro}, {Cisewski}, {Freeman}, {Genovese},
  {Lei}, {O'Connell}  \& {Wasserman}}{{Ciollaro} et~al.}{2014}]{Ciollaro2014}
{Ciollaro} M.,  {Cisewski} J.,  {Freeman} P.,  {Genovese} C.,  {Lei} J.,
  {O'Connell} R.,   {Wasserman} L.,  2014, arXiv e-prints, \href
  {https://ui.adsabs.harvard.edu/abs/2014arXiv1404.3168C} {p. arXiv:1404.3168}

\bibitem[\protect\citeauthoryear{{Davies} et~al.,}{{Davies}
  et~al.}{2018}]{Davies2018}
{Davies} F.~B.,  et~al., 2018, \mn@doi [\apj] {10.3847/1538-4357/aad7f8}, \href
  {https://ui.adsabs.harvard.edu/abs/2018ApJ...864..143D} {864, 143}

\bibitem[\protect\citeauthoryear{Dzuba, Flambaum  \& Webb}{Dzuba
  et~al.}{1999a}]{Dzuba1999a}
Dzuba V.~A.,  Flambaum V.~V.,   Webb J.~K.,  1999a, \mn@doi [Phys. Rev. A]
  {10.1103/PhysRevA.59.230}, 59, 230

\bibitem[\protect\citeauthoryear{Dzuba, Flambaum  \& Webb}{Dzuba
  et~al.}{1999b}]{Dzuba1999b}
Dzuba V.~A.,  Flambaum V.~V.,   Webb J.~K.,  1999b, \mn@doi [Phys. Rev. Lett.]
  {10.1103/PhysRevLett.82.888}, 82, 888

\bibitem[\protect\citeauthoryear{{Flambaum} \& {Shuryak}}{{Flambaum} \&
  {Shuryak}}{2008}]{flambaum08}
{Flambaum} V.~V.,  {Shuryak} E.~V.,  2008, in {Danielewicz} P.,  {Piecuch} P.,
   {Zelevinsky} V.,  eds,  American Institute of Physics Conference Series Vol.
  995, Nuclei and Mesoscopic Physic - WNMP 2007. pp 1--11 (\mn@eprint {}
  {physics/0701220}), \mn@doi{10.1063/1.2915601}

\bibitem[\protect\citeauthoryear{{Gill}, {Murray}  \& {Wright}}{{Gill}
  et~al.}{1981}]{GMW81}
{Gill} P.~E.,  {Murray} W.,   {Wright} M.~H.,  1981, {Practical optimization}.
London: Academic Press

\bibitem[\protect\citeauthoryear{{Hu} et~al.,}{{Hu} et~al.}{2019}]{Hu2019}
{Hu} J.,  et~al., 2019, \mn@doi [\mnras] {10.1093/mnras/stz739}, \href
  {https://ui.adsabs.harvard.edu/abs/2019MNRAS.485.5050H} {485, 5050}

\bibitem[\protect\citeauthoryear{{Hu} et~al.,}{{Hu} et~al.}{2021}]{Hu2021}
{Hu} J.,  et~al., 2021, \mn@doi [\mnras] {10.1093/mnras/staa3066}, \href
  {https://ui.adsabs.harvard.edu/abs/2021MNRAS.500.1466H} {500, 1466}

\bibitem[\protect\citeauthoryear{Kramida, {Yu.~Ralchenko}, Reader  \& {and NIST
  ASD Team}}{Kramida et~al.}{2022}]{NIST_ASD}
Kramida A.,  {Yu.~Ralchenko} Reader J.,   {and NIST ASD Team} 2022, {NIST
  Atomic Spectra Database (ver. 5.10), [Online]. Available:
  {\url{https://physics.nist.gov/asd}} [2022, November 21]. National Institute
  of Standards and Technology, Gaithersburg, MD.}

\bibitem[\protect\citeauthoryear{{Lee}, {Suzuki}  \& {Spergel}}{{Lee}
  et~al.}{2012}]{Lee2012}
{Lee} K.-G.,  {Suzuki} N.,   {Spergel} D.~N.,  2012, \mn@doi [\aj]
  {10.1088/0004-6256/143/2/51}, \href
  {https://ui.adsabs.harvard.edu/abs/2012AJ....143...51L} {143, 51}

\bibitem[\protect\citeauthoryear{{Lee}, {Webb}, {Carswell}  \&
  {Milakovi{\'c}}}{{Lee} et~al.}{2021a}]{Lee2020AI-VPFIT}
{Lee} C.-C.,  {Webb} J.~K.,  {Carswell} R.~F.,   {Milakovi{\'c}} D.,  2021a,
  \mn@doi [\mnras] {10.1093/mnras/stab977}, \href
  {https://ui.adsabs.harvard.edu/abs/2021MNRAS.504.1787L} {504, 1787}

\bibitem[\protect\citeauthoryear{{Lee}, {Webb}, {Milakovi{\'c}}  \&
  {Carswell}}{{Lee} et~al.}{2021b}]{Lee2021}
{Lee} C.-C.,  {Webb} J.~K.,  {Milakovi{\'c}} D.,   {Carswell} R.~F.,  2021b,
  \mn@doi [\mnras] {10.1093/mnras/stab2005}, \href
  {https://ui.adsabs.harvard.edu/abs/2021MNRAS.507...27L} {507, 27}

\bibitem[\protect\citeauthoryear{{Liske} et~al.,}{{Liske}
  et~al.}{2008}]{Liske2008}
{Liske} J.,  et~al., 2008, \mn@doi [\mnras] {10.1111/j.1365-2966.2008.13090.x},
  \href {https://ui.adsabs.harvard.edu/abs/2008MNRAS.386.1192L} {386, 1192}

\bibitem[\protect\citeauthoryear{{Magueijo}, {Barrow}  \& {Sandvik}}{{Magueijo}
  et~al.}{2002}]{magueijo02}
{Magueijo} J.,  {Barrow} J.~D.,   {Sandvik} H.~B.,  2002, \mn@doi [Physics
  Letters B] {10.1016/S0370-2693(02)02928-3}, \href
  {https://ui.adsabs.harvard.edu/abs/2002PhLB..549..284M} {549, 284}

\bibitem[\protect\citeauthoryear{Marconi et~al.,}{Marconi
  et~al.}{2016}]{Marconi2016}
Marconi A.,  et~al., 2016, in {Ground-based and Airborne Instrumentation for
  Astronomy VI, eds. Christopher J. Evans and Luc Simard and Hideki Takami}.
  SPIE, pp 676 -- 687, \mn@doi{10.1117/12.2231653}, \url
  {https://doi.org/10.1117/12.2231653}

\bibitem[\protect\citeauthoryear{Ong, Berengut  \& Flambaum}{Ong
  et~al.}{2013}]{Ong2013}
Ong A.,  Berengut J.~C.,   Flambaum V.~V.,  2013, \mn@doi [Phys. Rev. A]
  {10.1103/PhysRevA.88.052517}, 88, 052517

\bibitem[\protect\citeauthoryear{{Preval}, {Barstow}, {Holberg}  \&
  {Dickinson}}{{Preval} et~al.}{2013}]{Preval2013}
{Preval} S.~P.,  {Barstow} M.~A.,  {Holberg} J.~B.,   {Dickinson} N.~J.,  2013,
  \mn@doi [\mnras] {10.1093/mnras/stt1604}, \href
  {http://adsabs.harvard.edu/abs/2013MNRAS.436..659P} {436, 659}

\bibitem[\protect\citeauthoryear{{Raassen} \& {van Kleff}}{{Raassen} \& {van
  Kleff}}{1976}]{Raassen1976b}
{Raassen} A.~J.~J.,  {van Kleff} T. A.~M.,  1976, \mn@doi [Physica B+C]
  {10.1016/0378-4363(76)90112-1}, \href
  {https://ui.adsabs.harvard.edu/abs/1976PhyBC..85..180R} {85, 180}

\bibitem[\protect\citeauthoryear{{Raassen}, {van Kleff}  \& {Metsch}}{{Raassen}
  et~al.}{1976}]{Raassen1976a}
{Raassen} A.~J.~J.,  {van Kleff} T. A.~M.,   {Metsch} B.~C.,  1976, \mn@doi
  [Physica B+C] {10.1016/0378-4363(76)90015-2}, 84, 133

\bibitem[\protect\citeauthoryear{{S{\'a}nchez-Monge}, {Schilke}, {Ginsburg},
  {Cesaroni}  \& {Schmiedeke}}{{S{\'a}nchez-Monge} et~al.}{2018}]{Sanchez2018}
{S{\'a}nchez-Monge} {\'A}.,  {Schilke} P.,  {Ginsburg} A.,  {Cesaroni} R.,
  {Schmiedeke} A.,  2018, \mn@doi [\aap] {10.1051/0004-6361/201730425}, \href
  {https://ui.adsabs.harvard.edu/abs/2018A&A...609A.101S} {609, A101}

\bibitem[\protect\citeauthoryear{{Sandage}}{{Sandage}}{1962}]{Sandage1962}
{Sandage} A.,  1962, \mn@doi [Astrophysical Journal] {10.1086/147385}, \href
  {https://ui.adsabs.harvard.edu/abs/1962ApJ...136..319S} {136, 319}

\bibitem[\protect\citeauthoryear{{Suzuki}, {Tytler}, {Kirkman}, {O'Meara}  \&
  {Lubin}}{{Suzuki} et~al.}{2005}]{Suzuki2005}
{Suzuki} N.,  {Tytler} D.,  {Kirkman} D.,  {O'Meara} J.~M.,   {Lubin} D.,
  2005, \mn@doi [\apj] {10.1086/426062}, \href
  {https://ui.adsabs.harvard.edu/abs/2005ApJ...618..592S} {618, 592}

\bibitem[\protect\citeauthoryear{{Tamai}, {Koehler}, {Cirasuolo},
  {Biancat-Marchet}, {Tuti}  \& {Gonz{\'a}les Herrera}}{{Tamai}
  et~al.}{2018}]{ELT2018}
{Tamai} R.,  {Koehler} B.,  {Cirasuolo} M.,  {Biancat-Marchet} F.,  {Tuti} M.,
   {Gonz{\'a}les Herrera} J.~C.,  2018, in {Marshall} H.~K.,  {Spyromilio} J.,
  eds,  Society of Photo-Optical Instrumentation Engineers (SPIE) Conference
  Series Vol. 10700, Ground-based and Airborne Telescopes VII. p. 1070014,
  \mn@doi{10.1117/12.2309515}

\bibitem[\protect\citeauthoryear{{Uzan}}{{Uzan}}{2011}]{Uzan2011}
{Uzan} J.-P.,  2011, \mn@doi [Living Reviews in Relativity]
  {10.12942/lrr-2011-2}, \href
  {https://ui.adsabs.harvard.edu/abs/2011LRR....14....2U} {14, 2}

\bibitem[\protect\citeauthoryear{{Ward}, {Raassen}, {Kramida}  \&
  {Nave}}{{Ward} et~al.}{2019}]{Ward2019}
{Ward} J.~W.,  {Raassen} A.~J.~J.,  {Kramida} A.,   {Nave} G.,  2019, \mn@doi
  [\apjs] {10.3847/1538-4365/ab4ea3}, \href
  {https://ui.adsabs.harvard.edu/abs/2019ApJS..245...22W} {245, 22}

\bibitem[\protect\citeauthoryear{{Webb}, {Carswell}  \& {Lee}}{{Webb}
  et~al.}{2021}]{WebbVPFIT2021}
{Webb} J.~K.,  {Carswell} R.~F.,   {Lee} C.-C.,  2021, \mn@doi [\mnras]
  {10.1093/mnras/stab2895}, \href
  {https://ui.adsabs.harvard.edu/abs/2021MNRAS.508.3620W} {508, 3620}

\makeatother
\end{thebibliography}

\appendix
\newpage

\section{Full spectrum} \label{appendix}

\begin{figure*}
\centering
\includegraphics[width=0.95\linewidth]{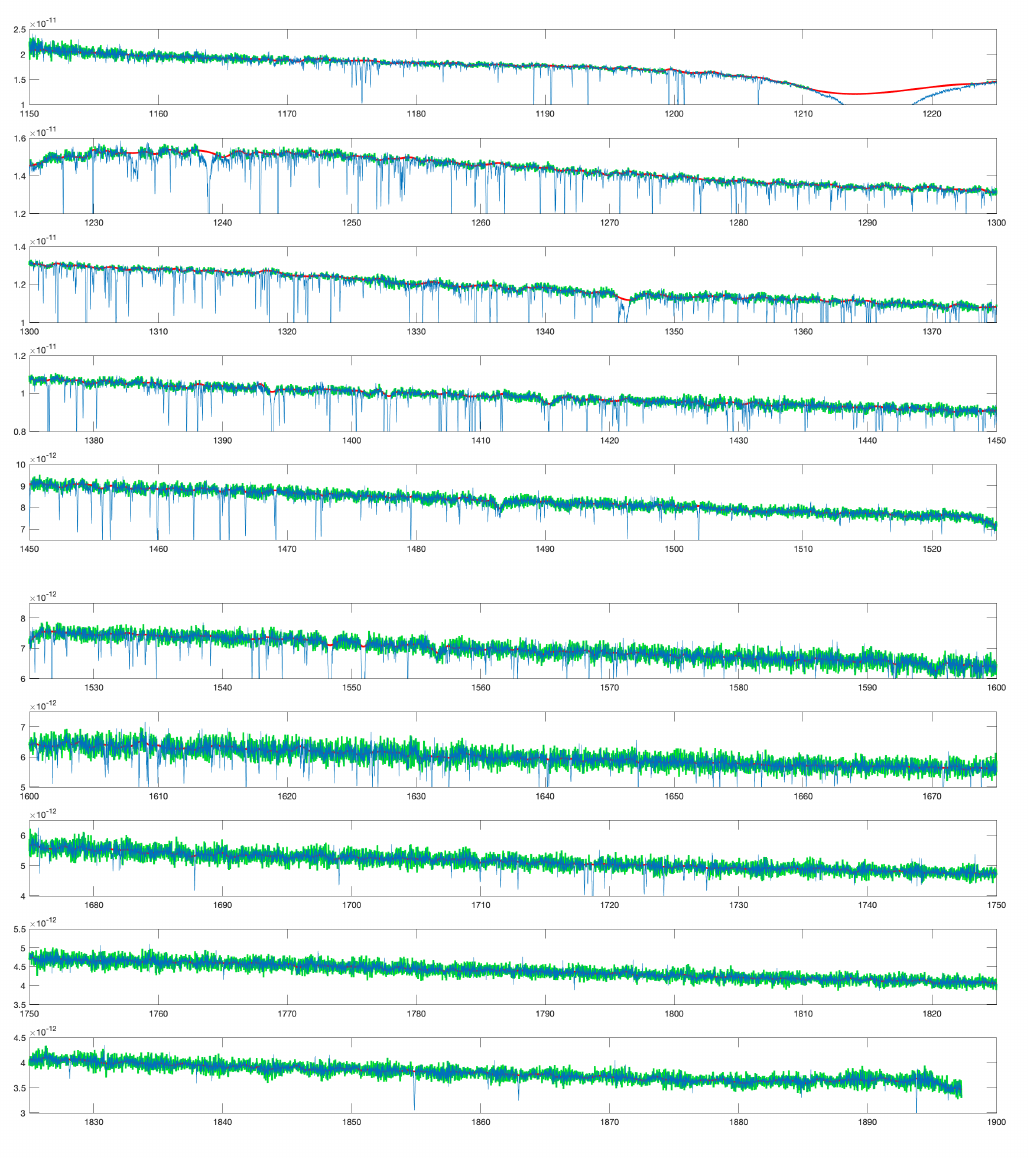}
\caption{Full spectrum, continuum model with 0.5 {\AA} knot spacing.
\label{fig:full0.5}
}
\end{figure*}

\begin{figure*}
\centering
\includegraphics[width=0.95\linewidth]{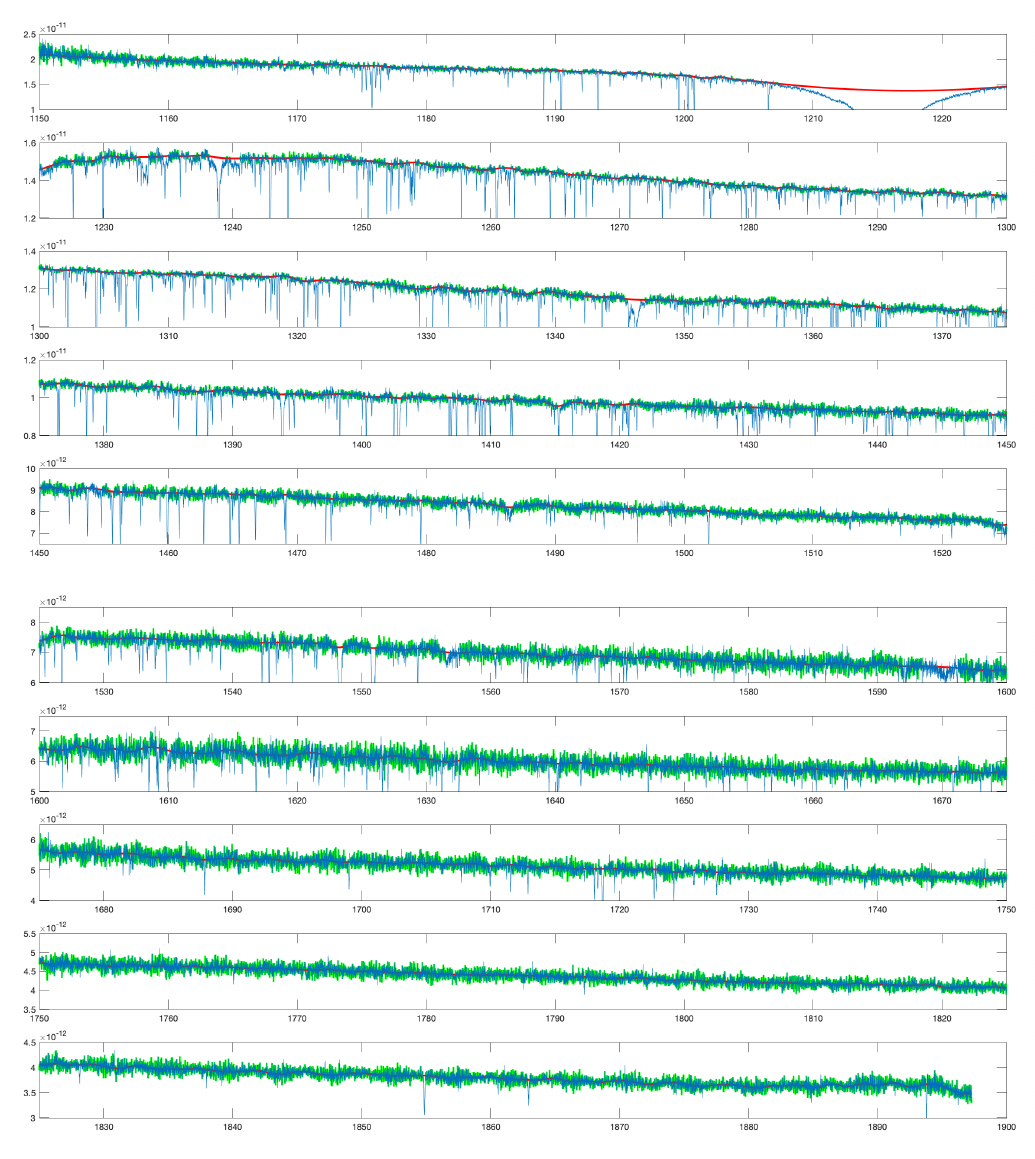}
\caption{Full spectrum, continuum model with 1.2 {\AA} knot spacing.
\label{fig:full1.2}
}
\end{figure*}

\bsp	
\label{lastpage}

\end{document}